\RequirePackage{arydshln}
\documentclass[aps,twocolumn,nofootinbib,superscriptaddress,preprintnumbers,10pt,floatfix]{revtex4-1}

\usepackage[utf8]{inputenc}

\usepackage{float}
\usepackage{amsmath,amssymb}
\usepackage{dsfont} 
\usepackage{hyperref}
\usepackage{graphicx}
\usepackage{enumitem}
\usepackage{mathtools}
\usepackage{bbold}
\usepackage{multirow}
\usepackage{ytableau}
\usepackage{youngtab}
\usepackage{braket}
\usepackage{soul}
\usepackage{cancel}
\usepackage{xcolor}
\usepackage[normalem]{ulem}

\usepackage{lipsum}

\usepackage{colortbl} 
\definecolor{Gray}{gray}{0.95}
\definecolor{RGray}{gray}{0.90}
\definecolor{CGray}{gray}{0.92}
\definecolor{nicegreen}{rgb}{0.1,0.5,0.1}

\usepackage{arydshln}

\newcommand{\mrm}[1]{\mathrm{#1}}
\newcommand{\re}[0]{\mrm{Re}}

\newcommand{\msbar}{{\overline{\rm{MS}}}}
\newcommand{\mev}{\mathrm{MeV}}
\newcommand{\gev}{\mathrm{GeV}}

\newcommand{\beq}{\begin{eqnarray}}
\newcommand{\eeq}{\end{eqnarray}}

\newcommand{\real}{{\sf I}\kern-.12em{\sf R}}
\newcommand{\comp}{{\sf I}\kern-.50em{\sf C}}
\newcommand{\unity}{{\sf I}\kern-.54em{\sf 1}}

\makeatletter
\g@addto@macro\bfseries{\boldmath}
\makeatother

\makeatletter
\renewcommand\paragraph{\@startsection{paragraph}{4}{\z@}%
                                    {3.25ex \@plus1ex \@minus.2ex}%
                                    {-1em}%
                                    {\normalfont\normalsize\bfseries}}
\makeatother

\allowdisplaybreaks
\begin{document}

\preprint{}
\preprint{}

\title{ Interpreting the results on exclusive $c\rightarrow s \mu \nu$ modes }

\author{D.~Be\v{c}irevi\'c}
\email{damir.becirevic@ijclab.in2p3.fr}
\affiliation{IJCLab, P\^ole Th\'eorie (Bat.~210), CNRS/IN2P3 et Universit\'e Paris-Saclay, 91405 Orsay, France}
\affiliation{CERN, Theory Department, 1211 Geneva 23, Switzerland}
\author{M.~Martines}
\email{matheus.martines.silva@usp.br}
\affiliation{Instituto de F\'isica, Universidade de S\~ao Paulo, R.~do Mat\~ao 1371, 05508-090 S\~ao Paulo, Brazil}
\author{S.~Rosauro-Alcaraz}
\email{salvador.rosauro@clermont.in2p3.fr}
\affiliation{Laboratoire de Physique de Clermont Auvergne (UMR 6533), 
CNRS/IN2P3 et Univ. Clermont Auvergne, 4 Av. Blaise Pascal, 63178 Aubi\`ere Cedex, France}
\author{O.~Sumensari}
\email{olcyr.sumensari@ijclab.in2p3.fr}
\affiliation{IJCLab, P\^ole Th\'eorie (Bat.~210), CNRS/IN2P3 et Universit\'e Paris-Saclay, 91405 Orsay, France}

\begin{abstract}
\vspace{5mm}
We examine the $q^2$-binned distributions of the $D\to K\mu\nu$ decay rate and of the corresponding forward-backward asymmetry which were recently measured by BESIII, showing a mild deviation 
from the Standard Model predictions.  We point out that the proposed solution to remedy the discrepancy by turning on a complex-valued New Physics coupling is in tension with the constraints  deduced from the LHC bounds on the high-$p_T$ tail of the relevant Drell-Yan process. We then show that there are 
several plausible scenarios that are compatible with both the measured low-energy and high-energy constraints but the selected couplings appear to be too small to 
be observed in the measurements of integrated observables, except for possibly the $q^2$-binned distribution of the angular observables relevant to $D_s\to \phi (\to KK) \mu\nu$ or 
$\Lambda_c\to \Lambda (\to p \pi) \mu\nu$ modes.   
\vspace{3mm}
\end{abstract}

\maketitle

\allowdisplaybreaks

\section{Introduction}\label{sec:intro}
The exclusive $c\to s\ell \nu$ decay modes are becoming increasingly interesting phenomenologically because of their potential of revealing (or constraining) 
the effects of physics beyond the Standard Model (BSM). The experimental precision achieved by the BESIII experiment for several exclusive modes is already 
such that it becomes possible to study the differential 
decay rates with very narrow $q^2$-bins.

Very recently, the BESIII Collaboration published their results for both $D\to Ke\nu$ and $D\to K\mu \nu$ modes~\cite{BESIII:2026ydr}. 
They provided the binned $q^2$-distribution of the decay width and of the forward-backward asymmetry, both for the charged and the neutral $D$-meson semileptonic decays. 
From their analysis, they conclude that the results for the muon in the final state deviate from 
the Standard Model (SM) by almost $2\sigma$, and in a follow-up 
paper they interpret such a deviation in terms of a non-zero complex scalar coefficient~\cite{BESIII:2026uin}.

In this letter, we will focus on the case of the muon 
in the final state and show that the solution proposed in Ref.~\cite{BESIII:2026uin} is not compatible with the bounds on that couplings derived from experimental LHC studies 
of the high-$p_T$ tails of $pp\to \mu \nu$ (+ soft jets)~\cite{Allwicher:2022gkm,Allwicher:2022mcg} (see also Ref.~\cite{Fuentes-Martin:2020lea}). We will then consider the couplings to New Physics (NP) to be real numbers and point out the scenarios in which two NP couplings are turned on to ensure the compatibility with current experimental data from low- and high-energy observables. In particular, we show that the right-handed vector current contribution can be conveniently constrained by the high-$p_T$ limits from $pp\to Wh$~\cite{Eboli:2025vks,Alioli:2017ces}. The combination of these LHC processes thus allows us to constrain the full set of operators entering charged currents at low energies. Finally, we will briefly discuss the future prospects for the High-Luminosity LHC (HL-LHC) for both mono-lepton and $Wh$ production.

\section{Low energy effective theory}\label{sec:eft}
Let us first remind the reader of the low-energy effective field theory (LEFT) that describes the $c\to s\mu \nu$ decays and captures generic dimension-six BSM operators. The effective Lagrangian can be written as:
\begin{align}
\label{eq:left}
    \mathcal{L}&_\mathrm{eff} = -2\sqrt{2}G_F V_{cs}\Big{[}(1+g_{V_L}^{cs\,\mu})\,\big{(}\bar{c}_{L}\gamma_\mu s_{L} \big{)}\big{(}\bar{\mu}_L \gamma^\mu\nu_{L}\big{)}\nonumber\\
    & +g_{V_R}^{cs\,\mu}\,\big{(}\bar{c}_{R}\gamma_\mu s_{R} \big{)}\big{(}\bar{\mu}_L \gamma^\mu\nu_{L}\big{)}
    +g_{S_L}^{cs\,\mu}\,\big{(}\bar{c}_{R} s_{L} \big{)}\big{(}\bar{\mu}_R \nu_{L}\big{)}\nonumber\\ 
    & +g_{S_R}^{cs\,\mu}\,\big{(}\bar{c}_{L} s_{R} \big{)}\big{(}\bar{\mu}_R\nu_{L}\big{)}
    +g_{T}^{cs\,\mu} \big{(}\bar{c}_{R}\sigma_{\mu\nu} s_{L} \big{)}\big{(}\bar{\mu}_R \sigma^{\mu\nu}\nu_{L}\big{)}\,\Big{]}\nonumber\\  
    & + \mathrm{h.c.}\,,
\end{align}
where $V_{cs}$ is the relevant Cabibbo--Kobayashi-Maskawa (CKM) matrix element, and $g_J^{cs\,\mu}$ are the couplings to the corresponding New Physics (NP) operators, 
while $J$ runs over vector $(V_{L(R)})$, scalar $(S_{L(R)})$, and tensor $(T)$ operators. To simplify our notation, we drop the flavor indices of the effective couplings, as we focus specifically on the $c\to s\mu\nu$ transition.
Setting all $g_J=0$, we recover the Fermi theory which is used to compute the decay rates of the leptonic and semileptonic decays of hadrons in the SM. From its comparison with experimental data one extracts the value of $V_{cs}$.~\footnote{Notice that in Eq.~\ref{eq:left} we do not consider couplings to a right handed neutrino, and in our analysis we neglect light neutrino masses.} 
In what follows, we use $\vert V_{cs}\vert =0.9735(1)$, as obtained by using CKM unitarity and the Cabibbo angle determined from (semi)leptonic kaon decays~\cite{Charles:2015gya,UTfit:2022hsi}.~\footnote{The quoted $|V_{cs}|$ value is consistent with the most recent updates of the CKM  unitarity triangle analysis, cf.~\url{http://www.utfit.org} and \url{http://ckmfitter.in2p3.fr}. Very recently a novel method was used to extract $|V_{cs}|$ from the inclusive semileptonic decay of $D_s$~\cite{DeSantis:2025yfm,DeSantis:2025qbb}, and the extracted value agrees with the one we use here.}

When introducing the BSM operators, it is desirable to use various decay modes and formulate a larger set of observables that can be obtained from the angular distributions 
of the semileptonic decays. If measured experimentally, a comparison of these observables with their values predicted in the SM 
would lead to independent constraints on $g_J$~\cite{Fajfer:2015ixa,Bolognani:2024cmr}. 
In Ref.~\cite{Becirevic:2020rzi} it was shown how this can be done in the case of pseudoscalar mesons. 
The advantage of that analysis was that most of the hadronic uncertainties were controlled 
by the latest lattice QCD (LQCD) results. The drawback, however, was a small number of observables, given that only the pseudoscalar mesons were considered. 
That situation can be significantly improved by including decays involving vector mesons in the final state, the angular analysis of which would offer more 
observables with different sensitivities to specific operators. 
For a long time, this opportunity was hindered by a comparably poor theoretical control over the corresponding hadronic matrix elements, since the LQCD determination of the relevant form factors was either inaccurate or inconclusive. Very recently, an important improvement in that respect has been 
made by the CLQCD collaboration~\cite{Fan:2025qgj}, the results of which we use when considering $D_s\to \phi \mu\nu$. The experimental results of the angular 
analysis of $D_s\to \phi (\to K^+K^-)\mu\nu$ have not yet been published.~\footnote{The $D_s\to \phi (\to K^+K^-)\mu\nu$ results from Ref.~\cite{BESIII:2023opt} are only provided in terms of the total branching ratio and the form-factor ratios at $q^2=0$, assuming the SM distributions. Differential results, analogous to those available for $D\to K\mu\nu$ decays, are necessary for a consistent theoretical reinterpretation within the SM and beyond.}  Instead, the full branching fraction was improved to~\cite{BESIII:2023opt}: 
\begin{align}\label{eqx:Dsphi}
\mathcal{B}(D_s\to \phi \mu \nu) = 2.25(11)\%\,. 
\end{align}   
However, this value leads to a rather weak constraint on the couplings $g_J$  due to the limited experimental precision, and it is therefore not relevant to the subject of this letter.~\footnote{It would be very beneficial for this research if the authors of Ref.~\cite{BESIII:2023opt} published even a subset of all the angular coefficients $\mathcal{I}_{1-9}$ appearing in Eqs.(7.1,7.2).}
It is worth mentioning that $D\to K^\ast \mu \nu$ will not be discussed here since a reliable LQCD determination of the corresponding hadronic form factors is not available. 
Finally, the angular distribution of $\Lambda_c\rightarrow \Lambda\left(\rightarrow p \pi\right)\mu^+\nu_\mu$ could be a rich source of observables too. 
Such an analysis has been partly made in Ref.~\cite{BESIII:2023jxv}, but the authors separated the $\Lambda_c\rightarrow \Lambda\mu \nu$ 
events from the $\Lambda_c\rightarrow \Lambda e \nu$ ones only for the branching ratio and the leptonic forward-backward asymmetry, and reported:
\begin{align}\label{eqx:LcLmu}
\mathcal{B}(\Lambda_c\to \Lambda \mu \nu) &= 3.48(17)\%\,,\cr
\langle A_\mathrm{fb}\rangle^{\Lambda_c\to \Lambda}_\mu  &= -0.22(4)\,.
\end{align}   
Thanks to the LQCD results of the relevant form factors~\cite{Meinel:2016dqj}, we can use the above experimental values and derive constraints on the couplings $g_J$, but they too turn out to be not sufficiently restrictive yet and therefore not useful for the subject of this letter. 

Finally, the modes that provide the strongest constraints for this analysis are:

\vspace*{5mm}
\noindent $\bullet$ The leptonic mode that is nowadays very precisely measured~\cite{BESIII:2023cym,BESIII:2024dvk}.  We take the PDG average~\cite{ParticleDataGroup:2024cfk}: 
\begin{align}\label{Dsmunu-exp}
\mathcal{B}(D_s\to \mu \nu)= 0.537(11)\%\,.
\end{align}   
Using Eq.~\eqref{eq:left} the expression for the decay rate reads:
\begin{align}
\label{eq:meson_leptonic_decay}
    \Gamma\left(D_s\rightarrow \mu\nu_{\ell}\right)=&S_\mathrm{EW}\frac{G_F^2|V_{cs}|^2f_{D_s}^2m_{D_s}m_\mu^2}{8\pi}\left(1-\frac{m_{\mu}^2}{m_{D_s}^2}\right)^2\nonumber\\
    &\times \left|1-g_A+g_P\frac{m_{D_s}^2}{m_{\mu}\left(m_c+m_s\right)}\right|^2\,,
\end{align}
from which the branching fraction is obtained as $\mathcal{B}\left(D_s\rightarrow \mu\nu_{\ell}\right)= \tau_{D_s} \, \Gamma\left(D_s\rightarrow \mu\nu_{\ell}\right)$, where 
$\tau_{D_s}$ is the $D_s$-meson lifetime. In the above expression, $S_\mathrm{EW}= 1.023$ accounts for the short-distance electro-weak correction~\cite{Sirlin:1977sv,Marciano:1993sh}, 
and we use $g_A \equiv g_{V_R}-g_{V_L}$, $g_P \equiv g_{S_R}-g_{S_L}$. Finally, $f_{D_s}$ is the decay constant that encodes the nonperturbative QCD effects of the hadronic matrix element:
\begin{align}
    \langle 0|\bar{s}\gamma^{\mu}\gamma_5 c|D_s(p)\rangle=i p^{\mu}f_{D_s}\,,
\end{align}
and it has been computed by means of numerical simulations of QCD on the lattice to a very high precision: 
$f_{D_s}=249.9(5)$~MeV~\cite{FlavourLatticeAveragingGroupFLAG:2024oxs}.  
Note that $g_P$ and the quark masses in Eq.~\eqref{eq:meson_leptonic_decay} are renormalization scheme and scale dependent. 
That dependence, of course, cancels out in the ratio $g_P/(m_c+m_s)$. In the following we will use $m_s^{\msbar}(2\,\gev) = 93.5(6)~\mev$, and $m_c/m_s=11.77(3)$~\cite{FlavourLatticeAveragingGroupFLAG:2024oxs}. 

\begin{table*}
\centering
\renewcommand{\arraystretch}{1.7}
\begin{tabular}{ | c | c | c   c  c | }
 \hline
 Quantity & Exp.~\cite{BESIII:2026ydr} & ETMC~\cite{Lubicz:2017syv} & FNAL/MILC~\cite{FermilabLattice:2022gku} & HTMC~\cite{Chakraborty:2021qav} \\
 \hline\hline
 $\mathcal{B}(D^0\to K^- \mu \nu)$ & 3.428(18)\%  & 3.4(2)\%  & 3.46(4)\% & 3.40(4)\% \\
 \hline
 $\mathcal{B}(D^+\to \bar K^0 \mu \nu)$ & 8.763(60)\%  & 8.7(6)\%  & 8.83(10)\% & 8.70(10)\% \\
 \hline
 $\langle A_\mathrm{fb}\rangle^{D^0\to K^-}_\mu$ & --5.88(30)\%  & --5.81(8)\%  & --5.57(3)\%  & --5.55(2)\%  \\
 \hline
 $\langle A_\mathrm{fb}\rangle^{D^+\to\overline K^0}_\mu$ & --5.44(37)\%  & --5.80(8)\%  & --5.56(3)\%  & --5.54(2)\%  \\
 \hline
\end{tabular}
\caption{\small \sl Comparison of the full branching fractions and the forward-backward asymmetry integrated over $q^2$. Experimental values are compared with the SM estimates by using the form factors as obtained in LQCD by three different collaborations.}
\label{tab:SM}
\end{table*}

\vspace*{5mm}
\noindent $\bullet$  The measurement of the semileptonic branching fractions of the $D$-meson decay to a kaon have been recently improved to~\cite{BESIII:2026ydr}:\footnote{Notice that the results of the two channels are fully compatible after taking into account the ratio of the mean lifetimes, $\tau_{D^+}/\tau_{D^0}=2.518(14)$~\cite{ParticleDataGroup:2024cfk}.} 
\begin{align}\label{eq:BRDK}
\mathcal{B}(D^0\to K^- \mu \nu)&= 3.429(18)\%\,, \cr
\mathcal{B}(D^+\to \bar K^0 \mu \nu) &= 8.763(60)\%\,,
\end{align}   
and the results of the integrated forward-backward asymmetry were reported as well~\cite{BESIII:2026ydr}:
\begin{align}\label{eq:AfbDK}
\langle A_\mathrm{fb}\rangle^{D^0\to K^-}_\mu  &= -5.88(30)\%\,, \cr 
 \langle A_\mathrm{fb}\rangle^{D^+\to\overline K^0}_\mu  &= -5.44(37)\%\,.
\end{align}

As we shall see below, the above experimental results are consistent with the SM predictions, but the $q^2$ distribution leads to a cumulative $\chi^2$ corresponding to nearly a $2\sigma$ deviation from the SM, independently of the set of form factors obtained from LQCD that is used. More specifically, using the Lagrangian given in Eq.~\eqref{eq:left}, the expression for the differential decay width reads: 
\begin{align}
    \label{eq:diff_BRmeson_semi-leptonic}
   & \frac{\mathrm{d} \Gamma (D\rightarrow K \mu\nu)}{\mathrm{d}q^2}=S_\mathrm{EW}\frac{G_F^2|V_{cs}|^2}{128\pi^3m_D^3}\, \sqrt{\lambda_{DK}}\, \left( 1- {m_\mu^2\over q^2}\right)^2\times  \cr
   &[f_+(q^2)]^2\Bigg\{ a_V(q^2) | 1+g_V|^2   + a_T(q^2) |g_T|^2  + a_S(q^2) |g_S|^2  \cr
& + a_{VT}(q^2) \re \left[  (1+g_V)g_T^\ast\right] + a_{VS}(q^2) \re \left[(1+g_V)g_S^\ast\right]  \Bigg\} ,
    \end{align}
where, for brevity, $\lambda_{DK} = \lambda(q^2,m_D^2,m_K^2)$, with $\lambda(a^2,b^2,c^2)=[a^2-(b-c)^2]\,[a^2-(b+c)^2]$, and 
 $g_V = g_{V_L}+g_{V_R}$, $g_S = g_{S_L}+g_{S_R}$. The remaining functions read: 
\begin{align}
&a_V(q^2) = {\lambda_{DK}\over 3}\left( 2+ \frac{m_\mu^2}{q^2}\right)+  \frac{m_\mu^2}{q^2} (m_D^2-m_K^2)^2 \left[ \frac{f_0(q^2)}{f_+(q^2)}\right]^2 ,\cr
&a_T(q^2) = {16\, q^2\, \lambda_{DK}\over 3 (m_D + m_K)^2}\left( 1+ \frac{2 m_\mu^2}{q^2}\right)\left[ \frac{f_T(q^2)}{f_+(q^2)}\right]^2  ,\cr
&a_S(q^2) = { q^2 \over (m_c - m_s)^2} (m_D^2-m_K^2)^2  \left[ \frac{f_0(q^2)}{f_+(q^2)}\right]^2  ,\cr
&a_{VT}(q^2) = {8 m_\mu \lambda_{DK}\over m_D + m_K}\, \frac{f_T(q^2)}{f_+(q^2)}\,  ,\cr
&a_{VS}(q^2) = {2 m_\mu \over m_c - m_s}\,(m_D^2-m_K^2)^2  \left[ \frac{f_0(q^2)}{f_+(q^2)}\right]^2 .
\end{align}
In deriving the above expression we used the standard decomposition of the hadronic matrix elements in terms of hadronic form factors:
\begin{align}
    \label{eq:DtoK_FF}
        \langle K(k)|\bar{s}\gamma^{\mu}c|D(p)\rangle = &\left[\left(p+k\right)^{\mu}-\frac{m_{D}^2-m_K^2}{q^2}q^\mu\right]f_+(q^2)\nonumber\\
        +&\frac{m_D^2-m_K^2}{q^2}q^{\mu}f_0(q^2)\,,\nonumber\\
        \langle K(k)|\bar{s}\sigma^{\mu\nu}c|D(p)\rangle  = &-i\left(p^\mu k^\nu-p^\nu k^\mu\right)\frac{2f_T(q^2)}{m_D+m_K}\,.
\end{align}
Note that the form factor $f_T(q^2)$ carries the renormalization scale dependence of the tensor density, which in the expression for the decay rate~\eqref{eq:diff_BRmeson_semi-leptonic} 
cancels the one in $g_T$. This dependence is not written explicitly for notational simplicity; however, unless stated otherwise, we use the $\msbar$ renormalization scheme with the scale $\mu = 2\,\gev$.~\footnote{The renormalization scale is chosen to match the one used in the calculation of the relevant hadronic matrix elements on the lattice~\cite{FlavourLatticeAveragingGroupFLAG:2024oxs,Gonzalez-Alonso:2017iyc}.}
The above form factors have been studied carefully in LQCD. Three collaborations have presented results obtained with gauge-field configurations with $N_{\mathrm{f}}=2+1+1$ dynamical quark flavors and multiple lattice spacings in order to control the extrapolation to the continuum limit. 
While FNAL/MILC~\cite{FermilabLattice:2022gku} and HPQCD~\cite{Chakraborty:2021qav,Parrott:2022rgu} used the staggered quark action, 
ETM Collaboration~\cite{Lubicz:2017syv,Lubicz:2018rfs} used the Wilson regularization with the (maximally) twisted mass term.

\begin{figure*}[htbp]
\centering
\includegraphics[scale=0.36]{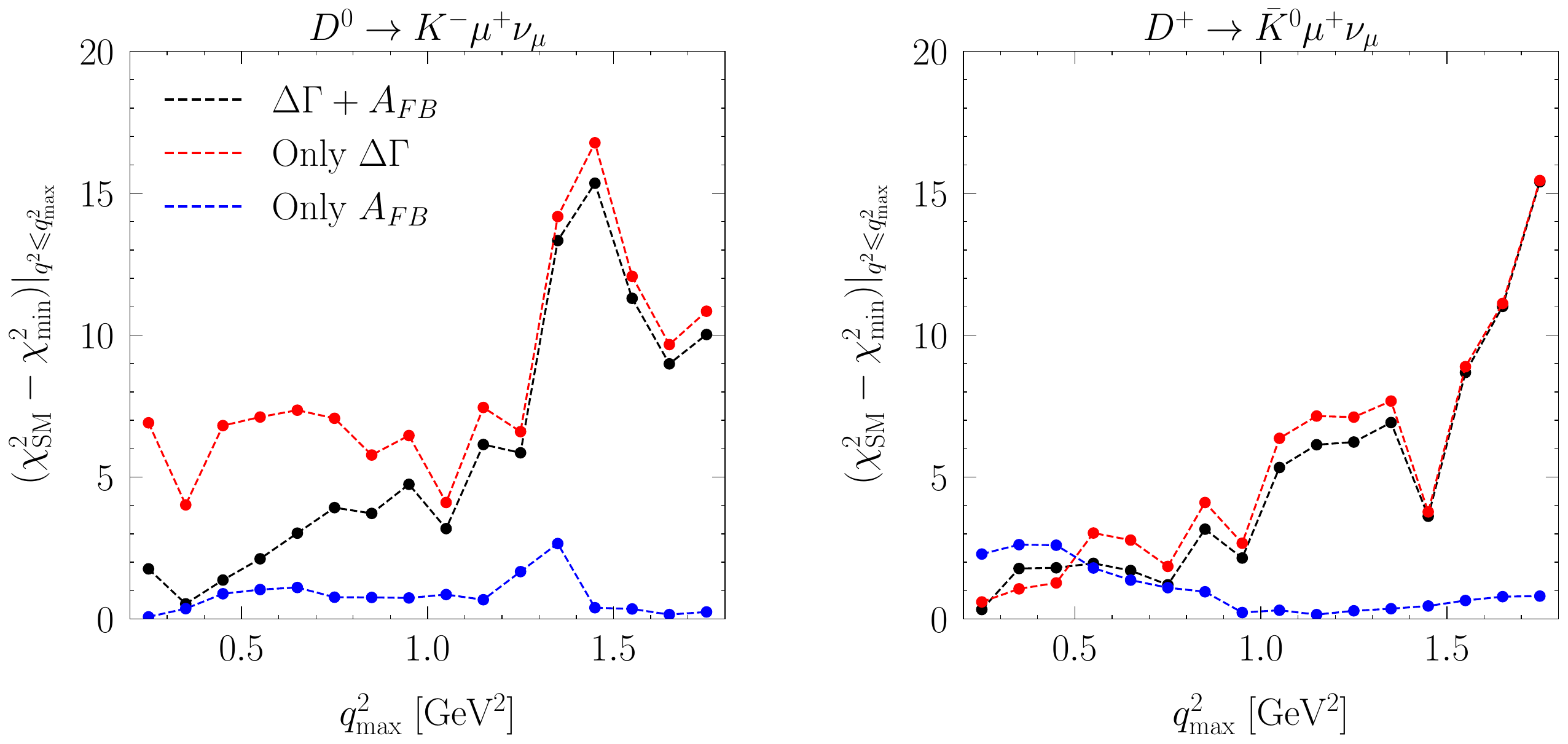}
\caption{\small\sl $\chi^2$-test showing the cumulative $\Delta \chi^2=\chi^2_\text{SM}-\chi^2_\mathrm{min}$ as one moves from lower to larger $q^2$-bins, obtained by using the experimental data from Ref.~\cite{BESIII:2026ydr} and the SM prediction for which the LQCD form factors from Ref.~\cite{Chakraborty:2021qav} have been employed. We see that in both modes the $q^2$-binned forward-backward asymmetry is not increasing $\chi^2$, and the main increase arises from the large $q^2$-bins of the differential decay rate. \label{fig:chi2}}
\end{figure*}

The integrated forward-backward asymmetry is defined as
\begin{align}
\langle A_\mathrm{fb}\rangle^{D\to K}_\mu = {1 \over \Gamma (D\rightarrow K \mu\nu)}\, \int_{m_\mu^2}^{q^2_\mathrm{max}} b(q^2)\, dq^2\,,
 \end{align}
where $q^2_\mathrm{max}=(m_D-m_K)^2$, and the function $b(q^2)$ in the numerator is given by the following expression~\cite{Becirevic:2020rzi}\footnote{Note that the angular convention adopted by the BESIII Collaboration in Refs.~\cite{BESIII:2026ydr, BESIII:2026uin} differs from that given in Ref.~\cite{Becirevic:2020rzi} by a factor of $\pi$, amouting to an overall minus sign for $A_{\mathrm{fb}}$ that we take into account in our numerical analysis.}: 
\begin{align}
    \label{eq:Afb_numerator}
   b(q^2)& =S_\mathrm{EW}\frac{G_F^2|V_{cs}|^2}{128\pi^3m_D^3}\, \lambda_{DK}\, \left( 1- {m_\mu^2\over q^2}\right)^2 m_\mu^2 \frac{m_D^2-m_K^2}{q^2}\times \cr
   &  f_+(q^2) f_0(q^2) \Bigg\{ |1+g_V|^2 + b_{VT}(q^2)  \re \left[(1+g_V)g_T^\ast\right] 
     \cr
   &  + b_{VS}(q^2)  \re \left[(1+g_V)g_S^\ast\right]  + b_{ST}(q^2)  \re \left[ g_S \, g_T^\ast\right]  \Bigg\},
     \end{align}
 with 
\begin{align}
&b_{VT}(q^2) = {4 q^2 \over m_\mu\, (m_D + m_K)}\, \frac{f_T(q^2)}{f_+(q^2)}\,  ,\cr
&b_{VS}(q^2) = {q^2 \over m_\mu \, (m_c - m_s)}\,,\cr
&b_{ST}(q^2) = {4q^2 \over m_\mu^2}{q^2\over  (m_c - m_s) (m_D+m_K)} \, \frac{f_T(q^2)}{f_+(q^2)} \, .
\end{align}
Note that, in measuring the binned forward-backward asymmetry, the experimentalists used:
\begin{align}
\left(A_\mathrm{fb}^\mathrm{bin}\right)^{D\to K}_\mu ={\displaystyle{ \int_{q_1^2}^{q_2^2} b(q^2)\, dq^2}\over \displaystyle{  \int_{q_1^2}^{q_2^2}  \frac{\mathrm{d}\Gamma (D\rightarrow K \mu\nu)}{\mathrm{d}q^2}\, dq^2} }\,,
 \end{align}
for a bin $q^2\in [q_1^2, q_2^2]$. We often refer to the binned partial decay rate as $\Delta \Gamma$. 
Using the LQCD results for the form factors, we compute the branching fraction and forward-backward asymmetries, and compare the resulting SM predictions with the experimental values in Tab.~\ref{tab:SM}. Note that the agreement with the SM is good for the different sets of form factors computed in LQCD. The deviation appears only when the finely binned $q^2$-distribution is considered. This is indeed shown in Fig.~\ref{fig:chi2}, where we show the $\chi^2$ test from which we see that the binned forward-backward asymmetry, for which the error bars are still large, is consistent with the SM, and that the main source of deviation with respect to the SM prediction occurs in the differential decay rate in the bins corresponding to larger values of $q^2$.

\section{Using the complex $g_S$}\label{sec:gs-im}

\begin{figure}[h]
\centering
\includegraphics[scale=0.4]{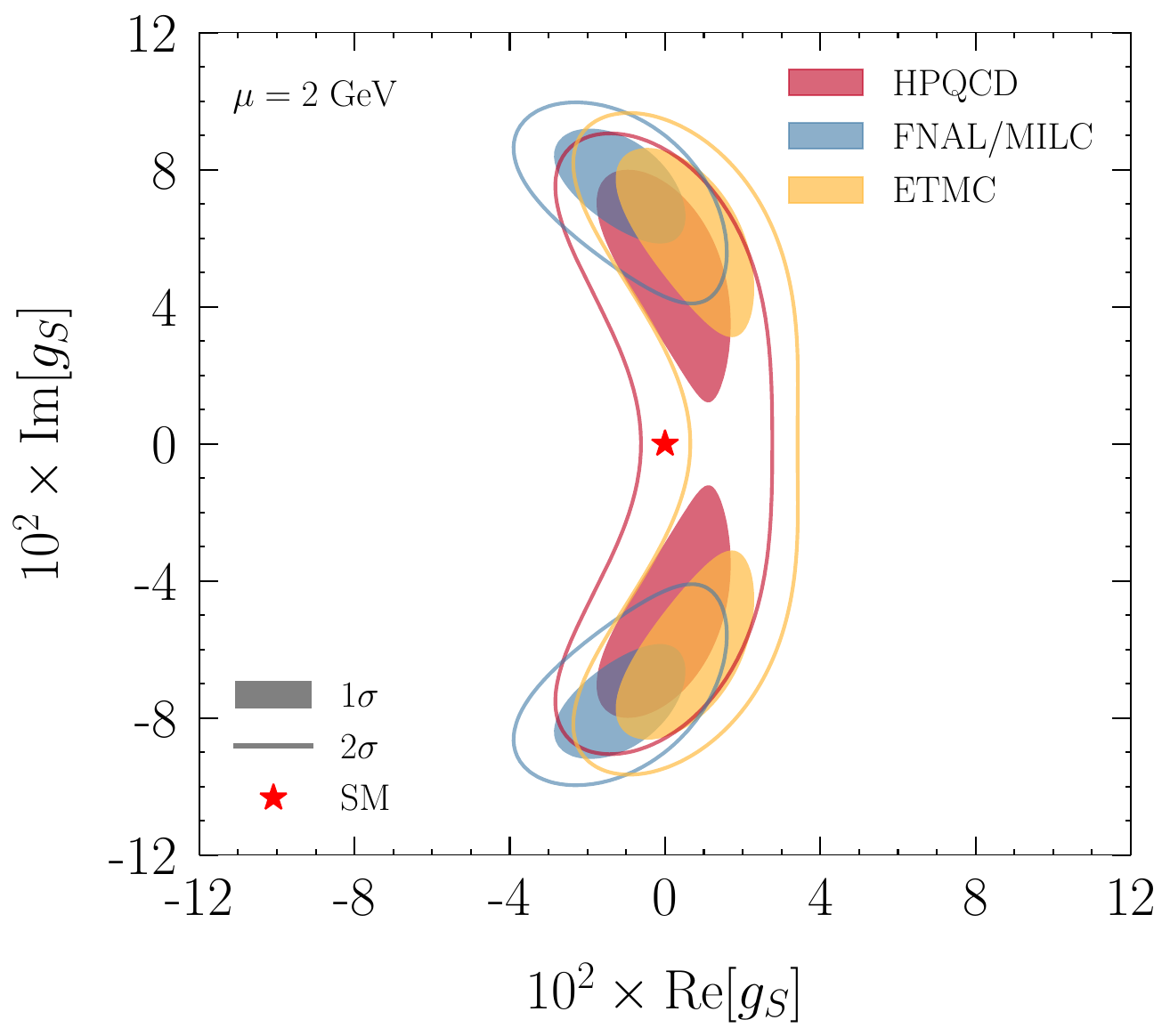}
\caption{\small\sl Accommodating the experimental data for the binned decay width and forward-backward asymmetry of $D\to K\mu \nu$~\cite{BESIII:2026ydr} by allowing $g_S = g_{S_L}+ g_{S_R}$ to be complex and non-zero. For that purpose the hadronic form factors obtained by three different LQCD Collaborations have been used: ETMC~\cite{Lubicz:2017syv}, FNAL/MILC~\cite{FermilabLattice:2022gku} and HPQCD~\cite{Chakraborty:2021qav}. \label{fig:1a}}
\end{figure}

As already mentioned in the introduction, after measuring the $q^2$-distribution of $\mathcal{B}(D\to K \mu\nu_\mu)$, $\mathcal{B}(D\to K e\nu_e)$, and of 
$\left(A_\mathrm{fb}\right)^{D\to K}_\mu$, $\left(A_\mathrm{fb}\right)^{D\to K}_e$, the BESIII collaboration assessed the significance of a possible scalar current contribution to the muon channel~\cite{BESIII:2026ydr,BESIII:2026uin}, finding a deviation from the SM prediction of almost $2\sigma$. Since their treatment of the form factors differs somewhat from ours, we examine their proposal using the form factors obtained by the three LQCD collaborations~\cite{FermilabLattice:2022gku,Chakraborty:2021qav,Parrott:2022rgu,Lubicz:2017syv,Lubicz:2018rfs}. The results are plotted in Fig.~\ref{fig:1a}.
Although the results of the various LQCD collaborations differ slightly, this difference cannot account for the deviation between theory and experiment; in fact, all three lattice results exclude compatibility with the SM at the $1\sigma$ level. In the following, to make the plots less crowded we will use the form factor results obtained by HPQCD~\cite{Chakraborty:2021qav,Parrott:2022rgu}.  
Since we consider the NP scale to be well above the electroweak scale, in addition to the low-energy constraints on the coupling $g_S$, one must also take into account the bounds derived from LHC studies of the high-$p_T$ tails of $pp\to \mu \nu$ (+ soft jets)~\cite{Allwicher:2022gkm,Allwicher:2022mcg} (see also Ref.~\cite{Fuentes-Martin:2020lea}), which constrain
\begin{align}
\qquad	|g_S| < 0.014\quad (95\%~\mathrm{CL})\,.
\end{align}

\begin{figure}[]
\centering
\includegraphics[scale=0.4]{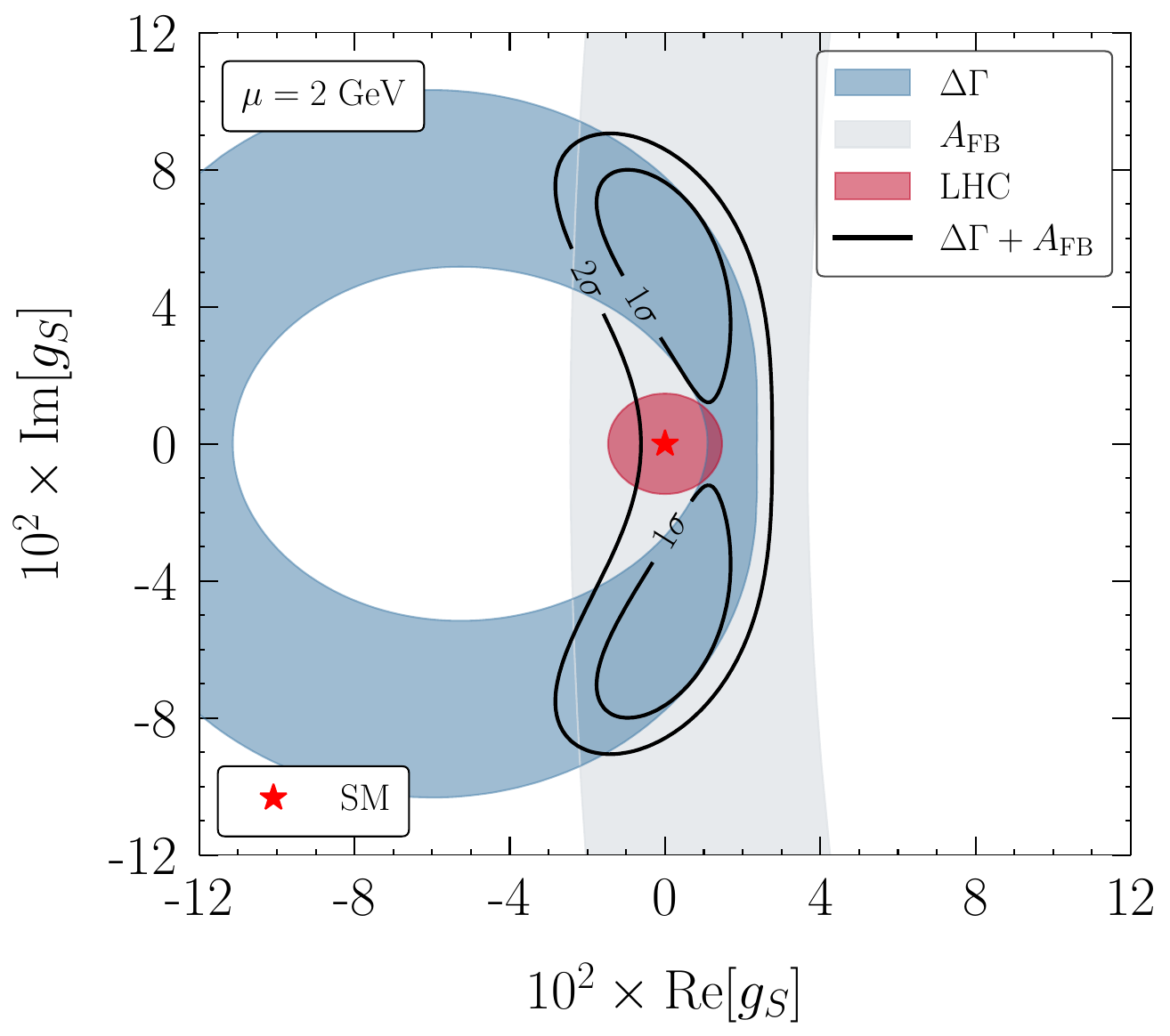}
\caption{\small\sl Low energy (gray and blue) and high energy (red) constraints on the complex coupling $g_S$ using the LQCD form factors from Ref.~\cite{Chakraborty:2021qav,Parrott:2022rgu}. See text. \label{fig:combine}}
\end{figure}

\noindent This result is also shown in Fig.~\ref{fig:combine}, from which we see that the low energy and high energy constraints are not compatible by almost $2\sigma$ and therefore a solution in terms of the complex valued coupling $g_S$ is not possible. Given the relevance of LHC bounds on constraining the NP couplings we consider, we devote the next section to thoroughly describe how they are obtained.

\section{SMEFT and LHC constraints}\label{sec:LHC}
The low-energy theory~\eqref{eq:left} can be matched with the so-called Standard Model Effective Field Theory (SMEFT)~\cite{Buchmuller:1985jz,Grzadkowski:2010es}, 
in which the leading gauge-invariant operators contributing to charged current interactions after electroweak symmetry breaking are of the type $\psi^4$ and $\psi^2 D H^2$, with $\psi$ 
representing a fermion, $D$ the covariant derivative and $H$ the Higgs doublet. The tree-level matching between the coefficients in Eq.~(\ref{eq:left}) and these SMEFT 
operators at $\mu\simeq m_W$ reads:~\footnote{We stress, once again, that we have dropped the flavor indices of the LEFT coefficients as we focus on the $c\to s\mu\nu$ transition, i.e., $g_J \equiv g_J^{cs\,\mu}$.}
\begin{align}
\label{eq:SMEFT_match}
    g_{S_R}^{cs\mu} &=-\frac{v^2}{2\Lambda^2}\sum_k\frac{V_{ck}}{V_{cs}}\mathcal{C}^*_{\substack{ledq\\\ell\ell2k}}\,,\\[0.35em]
    g_{S_L}^{cs\mu} &=-\frac{v^2}{2\Lambda^2}\frac{1}{ V_{cs}}\mathcal{C}_{\substack{lequ\\\ell\ell22}}^{(1)*}\,,\nonumber\\[0.35em]
    g_{V_L}^{cs\mu} &= -\frac{v^2}{\Lambda^2}\sum_k\frac{V_{ck}}{V_{cs}}\left[\mathcal{C}_{\substack{lq\\\ell\ell k2}}^{(3)}-\mathcal{C}^{(3)}_{\substack{Hq\\k2}}\right]+\frac{v^2}{\Lambda^2}\mathcal{C}_{\substack{Hl\\\ell\ell}}^{(3)}\,,\nonumber\\[0.35em]
    g_{V_R}^{cs\mu} & = \frac{v^2}{2\Lambda^2}\frac{1}{V_{cs}}\mathcal{C}_{\substack{Hud\\22}}\,,\nonumber\\[0.35em]
    g_{T}^{cs\mu} &= -\frac{v^2}{2\Lambda^2}\frac{1}{V_{cs}}\mathcal{C}^{(3)*}_{\substack{lequ\\\ell\ell22}}\,,\nonumber
\end{align}
where $v=(\sqrt{2} G_F)^{-1/2}$ is the Higgs vacuum expectation value, $\Lambda$ is the NP scale, and we use the operators defined, e.g.,~in Ref.~\cite{Allwicher:2022gkm} 
in the basis with diagonal down-quark Yukawa couplings. Flavor indices were explicitly written in the above equation to avoid confusion, but will be dropped in the following. The couplings are then run down to $\mu = 2\,\gev$. We obtain~\cite{Gonzalez-Alonso:2017iyc}:
\begin{align}
    \label{eq:running}
    g_{S_R}(2\,\mathrm{GeV})&\approx 1.72\, g_{S_R}(m_W)\,,\nonumber\\[0.35em]
    g_{S_L}(2\,\mathrm{GeV})&\approx 1.72\, g_{S_L}(m_W) - 0.02\, g_{T}(m_W)\,,\nonumber\\[0.35em]
    g_{V_{L(R)}}(2\,\mathrm{GeV})&\approx g_{V_{L(R)}}(m_W)\,,\nonumber\\[0.35em]
    g_{T}(2\,\mathrm{GeV})&\approx 0.82\, g_T^{\ell}(m_W)\,,
\end{align}
where we have considered three-loop QCD anomalous dimensions, and the (small) one-loop off-diagonal mixing induced by QED running. Furthermore, the running of the SMEFT 
coefficients from $\mu=1$~TeV down to $\mu=m_W$ for scalar and tensor coefficients amounts to~\cite{Gonzalez-Alonso:2017iyc}:
\begin{align}
    \mathcal{C}_{ledq}(m_W) &\approx 1.2\,\mathcal{C}_{ledq}(1\,\mathrm{TeV})\,,\nonumber\\*[0.35em]
    \mathcal{C}_{lequ}^{(1)}(m_W) &\approx 1.2\,\mathcal{C}^{(1)}_{lequ}(1\,\mathrm{TeV})-0.19\,\mathcal{C}^{(3)}_{lequ}(1\,\mathrm{TeV})\,,\nonumber\\*[0.35em]
    \mathcal{C}^{(3)}_{lequ}(m_W)&\approx 0.96\,\mathcal{C}^{(3)}_{lequ}(1\,\mathrm{TeV})\,,
\end{align}
where the flavor indices are omitted, and only the dominant running effects induced by QCD and electroweak interactions have been considered.

\begin{figure}[!t]
   \includegraphics[width=0.43\textwidth]{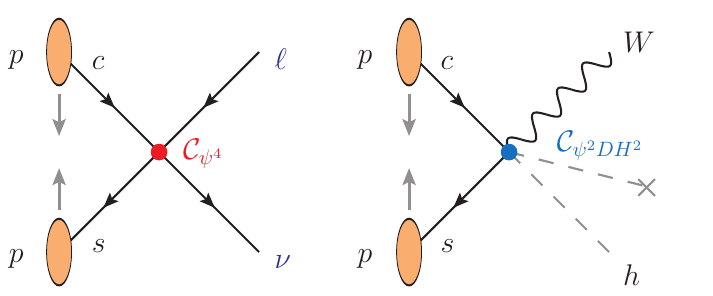}
    \caption{\small \sl Energy enhanced contribution from $\psi^4$ semileptonic operator to $pp\to\ell \nu$ (left panel) and from $\psi^2 DH^2$ Higgs-current operator to $pp\to W h$ (right panel).}
    \label{fig:lhc-charm}
\end{figure}

Note that the right-handed vector coupling $g_{V_R}$ is only generated through Higgs-current operators modifying the $W$ boson coupling to quarks, 
and is therefore lepton flavor universal.  It can be constrained through studies of $pp\rightarrow Wh$, which receive energy-enhanced contributions from these operators~\cite{Eboli:2025vks}. Instead, the four-fermion operators 
$\mathcal{O}_{ledq}$, $\mathcal{O}^{(1)}_{lequ}$, $\mathcal{O}_{lq}^{(3)}$, and $\mathcal{O}_{lequ}^{(3)}$ can be constrained from Drell-Yan processes, 
$pp\rightarrow \ell\nu$, as they induce a similar enhancement of the cross-section at high energies \cite{Allwicher:2022gkm,Allwicher:2022mcg}, cf.~Fig.~\ref{fig:lhc-charm}.~\footnote{By high energies here we refer to center-of-mass partonic energies $(\sqrt{\hat{s}})$ below the NP scale, 
such that $\sqrt{\hat{s}}<\Lambda$ and the EFT description of the processes is still consistent.}

Finally, the constraints on the couplings from the LHC data are derived from:
\begin{enumerate}
\item Mono-lepton Drell-Yan processes which directly probe the $\psi^4$ operators inducing charged-current interactions. At the parton level, these operators induce~\cite{Allwicher:2022gkm,Fuentes-Martin:2020lea}: 
\begin{align}    \label{eq:partonic_XS}
    &\dfrac{\hat{\sigma}\left(u_i\bar{d}_j\rightarrow \ell\nu\right)}{\hat{\sigma}\left(u_i\bar{d}_j\rightarrow \ell\nu\right)}_{\mathrm{SM}}=\bigg|1+g_{V_L}^{ij\,(H)}- \dfrac{\hat{s}}{m_W^2}g_{V_L}^{ij\ell\,(\psi)}\bigg|^2\cr
    &\qquad +\big|g_{V_R}^{\ij\,(H)}\big|^2  +\dfrac{\hat{s}^2}{m_W^4}\bigg[\frac{3}{4}\Big(|g_{S_R}^{ij\,\ell}|^2+|g_{S_L}^{ij\,\ell}|^2\Big) +4|g_T^{ij\,\ell}|^2\bigg]\,,\cr
\end{align}
where we have taken the limit $\hat{s}\gg m_W^2$ and considered a generic transition $u_i\bar{d}_j\rightarrow \ell\nu$, with flavor indices explicitly written again to avoid confusion. 
The contribution to the left-handed vector coupling $g_{V_L}^{ij\,\ell} \equiv g_{V_L}^{\ell\,(H)}+g_{V_L}^{ij\ell\,(\psi)}$ is divided into two contributions given by
\begin{align}
    g_{V_L}^{ij\,(H)}&=\frac{v^2}{\Lambda^2}\sum_k\frac{V_{ik}}{V_{ij}}\mathcal{C}^{(3)}_{\substack{Hq\\kj}}\,,\\*[0.35em]
    g_{V_L}^{ij\,\ell\,(\psi)}&=-\frac{v^2}{\Lambda^2}\sum_{k}\frac{V_{ik}}{V_{ij}}\mathcal{C}^{(3)}_{\substack{lq\\\ell\ell kj}}\,,\nonumber
\end{align}
as well as $\smash{g_{V_R}^{ij\,\ell} \equiv g_{V_R}^{ij\, (H)}}$. They separate the contributions modifying the $W$-boson couplings to quarks, which are lepton flavor universal, from the energy-enhanced one from contact interactions, respectively. 
The Higgs-current operator to leptons has been neglected given that it is tightly constrained by electroweak precision observables. We consider the data 
from ATLAS~\cite{ATLAS:2019lsy} as implemented in the package {\tt HighPT}~\cite{Allwicher:2022mcg}. 

\item Higgs $+$ $W$ boson production allows to directly constrain the Higgs-current operators. 
These operators induce an energy enhancement to the cross-section as~\cite{Eboli:2025vks}
\begin{align}
    \frac{\hat{\sigma}(u_i\bar{d}_j\rightarrow Wh)}{\hat{\sigma}(u_i\bar{d}_j\rightarrow Wh)_{\mathrm{SM}}}=
    \Bigg|1+\frac{g_{V_L}^{ij\,(H)}\hat{s}}{m_W^2}\Bigg|^2+\Bigg|\frac{g_{V_R}^{ij\,(H)}\hat{s}}{m_W^2}\Bigg|^2\,,
\end{align}
where only the energy-enhanced contributions from $d=6$ operators have been kept in the limit $\hat{s}\gg m_W^2$. Considering this data is essential to constrain $g_{V_L}$ in a model-independent way, as well as right-handed currents, since their components from $\psi^2 D H^2$ operators do not lead to energy-enhancement in mono-lepton production~\cite{Eboli:2025vks}. For our numerical analysis, we consider the ATLAS data on Higgs boson production in association with a $W$ boson, based on $140\,\mathrm{fb}^{-1}$~\cite{ATLAS:2024yzu} presented in the format of simplified template cross-sections (STXS)~\cite{LHCHiggsCrossSectionWorkingGroup:2016ypw}.
\end{enumerate}

\begin{figure*}[t]
\includegraphics[scale=0.55]{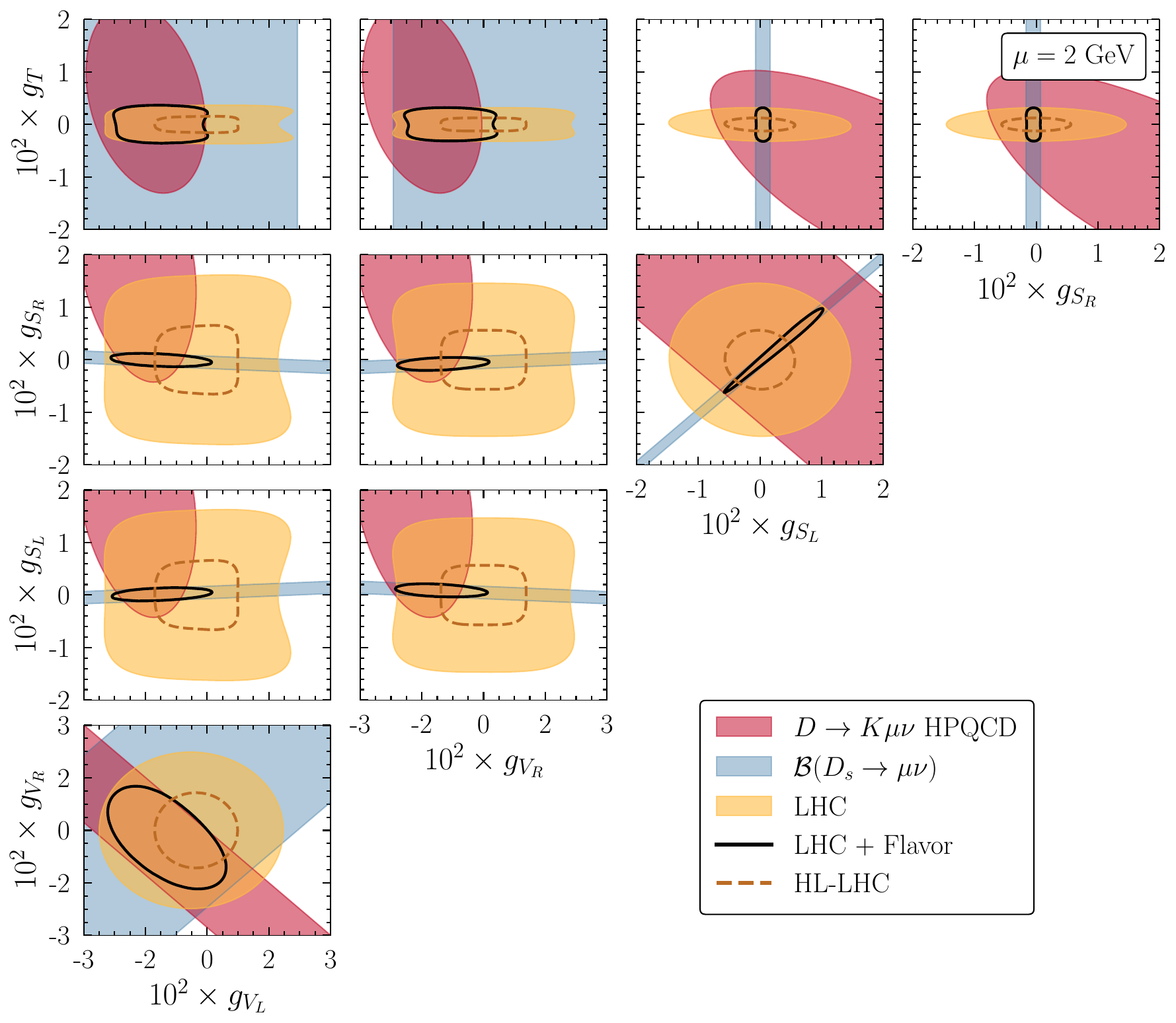}
\caption{\small \sl Low-energy and high-energy constraints on various combinations of (real valued) couplings appearing in Lagrangian~\eqref{eq:left}. Black curve in each plot correspond to the region of couplings that satisfy all the constraints to $2\sigma$. Dashed curve is the projection to the precision expected after completion of the High-Luminosity LHC, i.e. $3~\mathrm{ab}^{-1}$.}
\label{fig:2}
\end{figure*}

\section{Plausible scenarios}

In this section we use the constraints derived by using the decay rate and forward-backward asymmetry bin-by-bin, for which the BESIII provided experimental results with correlations~\cite{BESIII:2026ydr}, combined with the LQCD results for the $D\to K$ semileptonic form factor~\cite{Chakraborty:2021qav,Parrott:2022rgu}. We also use the leptonic decay result given in Eq.~\eqref{Dsmunu-exp}, combined with the precisely determined $f_{D_s}$ decay constant~\cite{FlavourLatticeAveragingGroupFLAG:2024oxs}, and consider the scenarios in which all the NP couplings in Eq.~\eqref{eq:left} are real valued.
Turning on a single NP coupling is not enough to ensure the compatibility with experimental data. The next simplest situation is to allow two NP couplings to be simultaneously 
non-zero. We show in Fig.~\ref{fig:2} all such scenarios where we also superpose the low energy constraints with the high-energy ones as discussed in the previous section.
We see that all such scenarios are acceptable to $2\sigma$.
Note however that the LHC constraints are quite strong and comprise the SM, which together with the tight constraint arising from the leptonic decay (also consistent with the SM prediction), 
only allow for tiny values of the NP couplings, cf.~regions within the black curves in Fig.~\ref{fig:2} to $2\sigma$. 
We find that these values are too small to generate any significant shift with respect to the SM values for any integrated observable mentioned so far, including the integrated observables that can be derived from the angular distribution of $D_s\to \phi (\to KK)\mu \nu_\mu$.  Instead, a binned experimental $q^2$-distribution of these observables ($q^2$-dependent coefficient in the angular $D_s\to \phi (\to KK)\mu \nu_\mu$ distribution) could provide valuable tests of the deviation observed in Ref.~\cite{BESIII:2026ydr}, and select among the scenarios presented in Fig.~\ref{fig:2}.

Finally, we note that data from the HL-LHC will further enhance the sensitivity of LHC constraints on charged-current transitions. The projected reach for an integrated luminosity of $3~\mathrm{ab}^{-1}$ is shown by the dashed curves in Fig.~\ref{fig:2} for both Drell-Yan \cite{Allwicher:2022gkm,Allwicher:2022mcg} and $Wh$ production~\cite{Eboli:2025vks}. This will provide the best sensitivity to all operators, with the exception of the pseudoscalar one, $g_P = g_{S_R} - g_{S_L}$, which is better constrained by the leptonic decays $D_s\to \mu\nu$, as $g_P$ lifts the helicity suppression of this process in the SM. Although we focus on the muon flavor in this letter, this conclusion also extends to operators with any other lepton. 

\section{Summary}

In this letter, we studied the recently published experimental results for the $D\to K\mu \nu$ decay~\cite{BESIII:2026ydr}. Although the total branching fraction and the integrated forward-backward asymmetry are consistent with the SM predictions, the detailed $q^2$-distribution (in narrow bins) reveals a deviation from the SM. We show that it is indeed so independently 
of the lattice QCD form factors used for such a comparison. In Ref.~\cite{BESIII:2026uin} it was proposed to accommodate that deviation by allowing a non-zero complex NP coupling to the scalar 
semileptonic operator. We show that such a solution, which leads to a non-zero $\mathrm{Im}(g_S)$ while $\mathrm{Re}(g_S)$ remains centered around zero, is not compatible with the LHC constraint 
on the same coupling arising from experimental studies of the high-$p_T$ tail of the corresponding Drell-Yan process. If, instead, we consider the complex $g_{S_L}$ coupling then we find 
that the constraints deduced from leptonic and semileptonic results are not compatible to each other. 
Combining the leptonic and semileptonic decays as constraints leads to many different scenarios in which any two of the couplings appearing in Eq.~\ref{eq:left} are allowed to be real and non-zero and are  compatible with the LHC constraints, cf. Fig.~\ref{fig:2}. These scenarios, however, are difficult to test elsewhere even though in the situations in which one considers a combination of couplings that involve $g_T$, the $q^2$-binned angular coefficients of $D_s \to \phi (\to KK) \mu \nu$, exhibit deviations with respect to the Standard Model that can be $\mathcal{O}(10\%)$ in a region of 
large $q^2$'s. These effects are dominated by the vector and tensor operator contributions, which turn out to be large for this type of transition. A more detailed study of these observables is left for future work.
\section{Acknowledgments}
\label{sec:akno}
 
This project has received support from the Agence Nationale de la Recherche under the contract
ANR-25-CE31-2504, from the European Union’s Horizon 2020 research and innovation programme under the Marie Skłodowska-Curie grant agreement N$^\circ$~860881-HIDDeN and N$^\circ$~101086085-ASYMMETRY, from the IN2P3 (CNRS) Master Project HighPTflavor, from the SPRINT/CNRS agreement supported by FAPESP under Contract No.~2023/00643-0 and from the USP-COFECUB project Uc Ph194-2. D.B.~thanks the CERN Department of Theoretical Physics for hospitality during this work. S.R.A.~thanks the Theory Department of IJCLab for their kind hospitality during this work. M.M.~acknowledges the support of Fundação de Amparo à Pesquisa do Estado de São Paulo (FAPESP) under the grant number 2022/11293-8.

\bibliography{biblio}

\end{document}